\documentclass[11pt]{article}
\begin{document}
\def \d {{\rm d}}
\def \A {{\rm A}}
\def \B {{\rm B}}
\def \CU {{\cal U}}
\def \CV {{\cal V}}

\title{The collision and snapping of cosmic strings generating
 spherical impulsive gravitational waves}

\author{J. Podolsk\'y\thanks{E--mail: {\tt Podolsky@mbox.troja.mff.cuni.cz}}
\\
\\ Institute of Theoretical Physics, Charles University,\\
V Hole\v{s}ovi\v{c}k\'ach 2, 18000 Prague 8, Czech Republic.\\
\\
and J. B. Griffiths\thanks{E--mail: {\tt J.B.Griffiths@Lboro.ac.uk}} \\ \\
Department of Mathematical Sciences, Loughborough University \\
Loughborough, Leics. LE11 3TU, U.K. \\ }

\date{\today}

\maketitle

\begin{abstract}
The Penrose method for constructing spherical impulsive gravitational waves is
investigated in detail, including alternative spatial sections and an arbitrary
cosmological constant. The resulting waves include those that are generated by
a snapping cosmic string. The method is used to construct an explicit
exact solution of Einstein's equations describing the collision of two
nonaligned cosmic strings in a Minkowski background which snap at their
point of collision.
\end{abstract}


\section{Introduction}

Several exact solutions of Einstein's equations have been published which
describe an impulsive spherical gravitational wave in a Minkowski background
generated, either by a snapping cosmic string (identified by a deficit angle),
or by an expanding string inside the sphere.

The first such solution was presented independently by both Gleiser and
Pullin~\cite{GlePul89} and Bi\v{c}\'ak and Schmidt~\cite{BicSch89}. The
first of these \cite{GlePul89} was obtained by pasting two appropriate forms of
Minkowski space either side of the spherical wavefront, while the second
\cite{BicSch89} was obtained as a limiting case of a solution with
boost-rotation
symmetry. In this case, two null particles recede from a common point
generating
an impulsive spherical gravitational wave and there are either cosmic strings
attaching each particle to infinity, or there is an expanding cosmic string
along the axis of symmetry separating the two particles. However, as pointed
out by Bi\v{c}\'ak~\cite{Bicak90}, this solution does not strictly describe a
snapping cosmic string, but two semi-infinite cosmic strings initially
approaching at the speed of light and then separating again at the instant at
which they collide.

In fact a general method for constructing expanding impulsive spherical
gravitational waves had previously been presented by Penrose \cite{Pen72}.
This involves cutting Minkowski space-time along a null cone and then
re-attaching the two pieces with a suitable ``warp''. The explicit general
solution written in a continuous coordinate system was given by Nutku and
Penrose~\cite{NutPen92} and Hogan \cite{Hogan93}, \cite{Hogan94}. The above
solutions describing ``snapping cosmic strings'' are included within this
family.

These solutions have been extended \cite{Hogan92} to include a
cosmological constant. With this they describe expanding spherical
gravitational waves in de~Sitter and anti-de~Sitter backgrounds. Further, the
general class of solutions of this type has been shown \cite{PodGri99}
to be equivalent to impulsive limits of the Robinson--Trautman type~N
class of solutions. Finally, it may be mentioned that Horta\c{c}su and
his colleagues \cite{Hort90}--\cite{BiHoOz96} have considered aspects
of particle creation in these backgrounds.

It may be noted that Hogan~\cite{Hogan95} has also considered
imploding-exploding gravitational waves. In this he has simply attached an
impulsive wave on a past null cone to one on a future null cone without
considering the necessary sources of the wave. Outside the cone, a snapping
string must evolve continuously. However, this construction does permit a
string inside an expanding cone to be different from one inside the prior
imploding cone, although the singular event at which the two cones join would
then require some physical justification.

The purpose of the present paper is to describe the Penrose method and its
interpretation in detail and in full generality, using alternative spatial
sections and including an arbitrary cosmological constant. This will enable us
to construct an exact solution which describes the gravitational wave that
would be generated by the collision (and consequent breaking) of a pair of
moving cosmic strings in a Minkowski background. Such a situation was outlined
in \cite{NutPen92}. The explicit solution is given in section~VI.

\section{The explicit Penrose construction}

As mentioned above, Penrose \cite{Pen72} has described a ``cut and paste''
method for constructing an expanding spherical gravitational wave in a
Minkowski background. In this method, appropriate junction conditions across
the null cone guarantee that the vacuum field equations are satisfied.
However, there is an impulsive component in the Weyl tensor representing an
impulsive gravitational wave located on this null cone.

The above procedure will now be performed explicitly. However, for
later convenience, we will derive a more general form of the solution for
an impulsive spherical wave in a Minkowski background than that outlined in
\cite{Pen72}. In addition, we will also include a cosmological
constant $\Lambda$, so that the resulting solutions will also describe
expanding spherical impulsive waves in de~Sitter and anti-de~Sitter
backgrounds. In this approach, it is convenient to start with the line
element for space-times of constant curvature in the manifestly conformally
flat form
 \begin{equation}
 \d s^2 = {2\d\eta\,\d\bar\eta  -2\d u\,\d
v\over
[1+{1\over6}\Lambda(\eta\bar\eta-uv)]^2} ,
 \label{nullMink}
 \end{equation}
 where the relation to conformal cartesian coordinates is
$\eta={1\over\sqrt2}(x+iy)$, $u={1\over\sqrt2}(t+z)$ and
$v={1\over\sqrt2}(t-z)$.

We may now perform the transformation
 \begin{eqnarray}
 v&=& {V\over p}-\epsilon U, \nonumber\\
 u&=& Z\bar Z \,{V\over p}-U,
 \label{inv} \\
 \eta&=& {V\over p}\,Z, \nonumber
 \end{eqnarray}
 where
 \begin{equation}
 p=1+\epsilon Z\bar Z, \qquad \epsilon=-1,0,1.
 \label{pandep}
 \end{equation}
The parameter $\epsilon$ is related to the Gaussian curvature of
the 2-surfaces given by $U=$~const., $V=$~const. \cite{PodGri99}.
With this, the metric (\ref{nullMink}) becomes
 \begin{equation}
\d s^2 = {{ \displaystyle
 2 {V^2\over p^2}\,\d Z\,\d\bar Z +2\d U\,\d V -2\epsilon\,\d U^2}\over
[1+{1\over6}\Lambda U(V-\epsilon U)]^2}.
 \label{U<0}
 \end{equation}

Let us also consider the alternative, and more involved, transformation
given by
 \begin{eqnarray}
 v&=& AV-DU,   \nonumber\\
 u&=& BV-EU,
 \label{transe3} \\
 \eta&=& CV-FU,  \nonumber
 \end{eqnarray}
where
 \begin{eqnarray}
&&A= \frac{1}{p|h'|},\qquad
B= \frac{|h|^2}{p|h'|},\qquad
C= \frac{h}{ p|h'|},   \nonumber\\
&&D= \frac{1}{|h'|}\left\{
\frac{p}{4} \left|\frac{h''}{h'}\right|^2+\epsilon
\left[1+\frac{Z}{2}\frac{h''}{h'}+\frac{\bar Z}{2}\frac{\bar h''}{\bar h'}
\right]\right\},\nonumber\\
&&E= \frac{|h|^2}{|h'|}
\bigg\{ \frac{p}{4}\left|\frac{h''}{h'}-2\frac{h'}{h}\right|^2
+\epsilon\left[ 1+\frac{Z}{2}
\left(\frac{h''}{h'}-2\frac{h'}{h}\right)+\frac{\bar Z}{2}
\left(\frac{\bar h''}{\bar h'}-2\frac{\bar h'}{\bar h}\right)
\right]\bigg\}\  ,\nonumber\\
&&F= \frac{h}{|h'|}\bigg\{
\frac{p}{4}\left(\frac{h''}{h'}-2\frac{h'}{h}\right)
\frac{\bar h''}{\bar h'}
+\epsilon\left[1+
 \frac{Z}{2}\left(\frac{h''}{h'}-2\frac{h'}{h}\right)
+\frac{\bar Z}{2}\frac{\bar h''}{\bar h'}\right]\bigg\}\ ,
 \label{transe4}
 \end{eqnarray}
 $h=h(Z)$ is an arbitrary holomorphic function (apart from its singular
regions), and the derivative with respect to $Z$ is denoted by a prime. With
this, the Minkowski, de~Sitter and anti-de~Sitter metric (\ref{nullMink})
becomes
 \begin{equation}
\d s^2 = {{ \displaystyle
2\left| {V\over p}\,\d Z+U\,p\,\bar H\,\d\bar Z \right|^2
+2\d U\,\d V -2\epsilon\,\d U^2} \over
[1+{1\over6}\Lambda U(V-\epsilon U)]^2} ,
 \label{U>0}
 \end{equation}
 where \ $2H(Z)=\beta'-{1\over2}\beta^2$, \ $\beta=\alpha'/\alpha$ \ and \
$\alpha=h'$. \ Thus,
 \begin{equation}
 H(Z)={1\over2} \left[{h'''\over h'}-{3\over2}\left({h''\over h'}\right)^2
\right],
 \label{Schwarz}
 \end{equation}
 is the Schwarzian derivative. Notice that the transformation (\ref{transe3})
and (\ref{transe4}) reduces to (\ref{inv}) when $h=Z$.

In the coordinates of both line elements (\ref{U<0}) and (\ref{U>0}), the null
hypersurface \ $U=0$ \ represents a null cone (an expanding sphere) \
\hbox{$\eta\bar\eta-uv=0$} \
in the background. Moreover, the reduced 2-metrics on this cone are identical.
Following Penrose's ``cut and paste'' method \cite{Pen72}, we may take the
line element (\ref{U<0}) for $U<0$ and attach this to (\ref{U>0}) for $U>0$.
The resulting line element can then be written in the combined form
 \begin{equation}
\d s^2 = {{ \displaystyle
 2\left| {V\over p}\,\d Z+U\Theta(U)\,p\,\bar H\,\d\bar Z \right|^2
+2\d U\,\d V -2\epsilon\,\d U^2 }
\over [1+{1\over6}\Lambda U(V-\epsilon U)]^2},
 \label{Ln0en0}
 \end{equation}
where $\Theta(U)$ is the Heaviside step function. This combined metric,
which was presented for a Minkowski background in
\cite{NutPen92}-\cite{Hogan94}, with a cosmological constant in
\cite{Hogan92}, and in the most general form in \cite{PodGri99}, is
explicitly continuous everywhere, including across the null
hypersurface $U=0$. However, the discontinuity in the derivatives of the
metric yields impulsive components in the curvature tensor proportional
to the Dirac $\delta$-function.

It may be observed that the Penrose junction conditions can be obtained by
comparing the transformations (\ref{inv}) at $U=0_-$ with
(\ref{transe3}--\ref{transe4}) at $U=0_+$. This gives the identification
 \begin{eqnarray}
 && (Z, \bar Z, V, U=0)_{M^-} =\left(h(Z), \bar h(\bar Z),
{1+\epsilon h\bar h\over1+\epsilon Z\bar Z}{V\over|h'|}, U=0 \right)_{M^+},
 \label{Junct}
 \end{eqnarray}
 where the space-time has been divided into two halves
$M^{\scriptscriptstyle-}(U<0)$ inside the null cone, and
$M^{\scriptscriptstyle+}(U>0)$ outside.

Labelling coordinates $(x^1,x^2,x^3,x^4)=(Z,\bar Z,V,U)$, and using the tetrad
 \begin{eqnarray}
 k^\mu &=& \left[1+{\textstyle{1\over6}}\Lambda U(V-\epsilon U)\right]^2
\delta^\mu_3, \nonumber\\
 \ell^\mu &=& -\delta^\mu_4 -\epsilon\,\delta^\mu_3, \nonumber\\
 m^\mu &=& p^2 {1+{1\over6}\Lambda U(V-\epsilon U) \over
V^2-p^4U^2\Theta(U)H\bar H} \left( {V\over p}\,\delta^\mu_2
-pU\Theta(U)\bar H\,\delta^\mu_1 \right), \nonumber
 \end{eqnarray}
 the non-zero components of the curvature tensor for the line element
(\ref{Ln0en0}) can be shown to be given by
 \begin{equation}
 \Psi_4={p^2H\over V}\,\delta(U), \qquad
\Phi_{22}={p^4H\bar H\over V^2}\,U\,\delta(U).
 \label{Psi4}
 \end{equation}
 This indicates an impulsive gravitational wave component. It also confirms
that the space-time is vacuum everywhere except on the wave surface at $V=0$
and at possible singularities of the function $p^2H(Z)$.

It may be
observed that the metric (\ref{U<0}) contains a coordinate
singularity at
$V=0$. However, this becomes a physical singularity on the wave
surface
$U=0$ in the above construction. Unfortunately, for $\epsilon=0$,
$V=0$ is
a singular null {\it line} on the surface $U=0$. In fact, as seen
from
(\ref{inv}), this is a common line ($x=y=0$, $z=t$) to all the null
cones
$U=$~const. Thus, for $\epsilon=0$, there is a singular line on the
wave
surface where $V=0$, in addition to possible singular points of $H$.
For a
physical interpretation of these solutions, it would be better to
remove the
singularity $V=0$ from the wavefront. This can be achieved by
considering
solutions with $\epsilon\ne0$.  For these cases, as observed by
Hogan~\cite{Hogan94}, the family of null cones $U=$~const. have vertices on a
timelike line ($x=y=z=0$) if $\epsilon=1$ and on a spacelike line
($x=y=t=0$) if $\epsilon=-1$. For either of these cases, the singularity at
$V=0$ appears only at the vertex of the null cone $x=y=z=t=0$, which may be
considered as the ``origin'' of the spherical wave.

\section{A geometrical description of the impulse}

Significantly, the construction of the spherical impulsive wave as described
above admits an interesting geometrical interpretation. Specifically, the
ratio
 \begin{equation}
 \xi \equiv{\eta\over v} ={x+iy\over t-z}
 \end{equation}
 is the well known relation for a stereographic (one-to-one) correspondence
between a sphere and an Argand plane by a projection from the North pole
onto a plane through the equator (see chapter~1 of \cite{PenRin84}). This
permits us to represent the wave surface $U=0$ (taken at a typical or
rescaled time $t=1$) either as a Riemann sphere, or as its associated
complex plane. Conversely, any point $\xi$ on the complex plane, taken as
$z=0$, identifies a unique point $P$ on the sphere with coordinates
 $$ x={\xi+\bar\xi\over1+\xi\bar\xi}, \qquad
y={i(\bar\xi-\xi)\over1+\xi\bar\xi}, \qquad
z={\xi\bar\xi-1\over\xi\bar\xi+1}. $$
 In terms of standard spherical coordinates,
$\xi=\cot{\theta\over2}\,e^{i\phi}$.

We may also recall some important properties of the stereographic projection.
Any circle on the Riemann sphere maps to a circle in the complex Argand
plane, and vice versa. As a special case, any circle passing through the
North pole of the Riemann sphere maps to a straight line in the complex plane.
A great circle which passes through the North pole maps to a straight
line
through the origin of the plane. We also note that the
stereographic
projection is conformal, i.e. angle preserving.

Returning to the impulsive spherical wave, we observe from (\ref{inv})
and
(\ref{transe3}--\ref{transe4}) that on $U=0$
 \begin{equation}
\xi=\cases{Z \qquad  &for \quad $U=0_-$ \cr
\noalign{\medskip}
 h(Z) \qquad
&for \quad $U=0_+$ \cr }
 \label{xi}
 \end{equation}
 This permits us to
represent the Penrose junction condition (\ref{Junct})
across the wave
surface as a mapping on the complex Argand plane
$Z\to h(Z)$. This is
equivalent to mapping points $P_-$ on the ``inside''
of the wave surface to
the identified points $P_+$ on the ``outside'', see
Fig.~\ref{Fig.1}.

Normally, we will assume that the ``inside'' represented
by $Z$ covers the complete sphere, but the function $h(Z)$ will not
generally cover the entire sphere on the ``outside''. The restrictions
on the range of the function $h(Z)$ together with its specific
character will correspond to particular physical situations as will be
described below.

For the sake of completeness, the results of these two sections
have been given with an arbitrary cosmological constant included.
However, for the remainder of this paper, we will only consider
the case when $\Lambda=0$.

\section{A snapping cosmic string}

The utility of the
above geometrical approach may be demonstrated by the simplest and
physically interesting case of an impulsive spherical wave generated by a
``snapping cosmic string''. Without loss of generality, the string
may be taken to be located along the $z$-axis perpendicular to the
above complex plane.

It may be noted that the initial Gleiser--Pullin~\cite{GlePul89}
solution describing this situation was presented in a slightly
different form to that given above. Essentially, it employed two
different constant non-zero values of $H$ in the regions inside
and outside the spherical wave. This is effectively equivalent
to taking $\xi=e^Z$ for $U<0$ and $\xi=e^{(1-\delta)Z}$ for
$U>0$, where $\delta$ is a real positive constant.

Here we will describe the same solution. However we will use the above
notation as also presented by Nutku and Penrose~\cite{NutPen92}.
Taking $\Lambda=0$, the line element (\ref{Ln0en0}) becomes

\begin{equation}
\d s^2 =
 2\left| {V\over p}\,\d Z+U\Theta(U)\,p\,\bar
H\,\d\bar Z \right|^2
+2\d U\,\d V -2\epsilon\,\d U^2 .
 \label{L=0en0}
\end{equation}
where $p$ and $\epsilon$ are given by
(\ref{pandep}).

According to the above construction (\ref{xi}), in the interior
region $U<0$, the complete spherical surface is covered by $\xi=Z$.
It is then appropriate to represent the outer region $U>0$ by
$\xi=h_1(Z)$ where
 \begin{equation}
 h_1(Z)=Z^{1-\delta}.
 \label{h1}
 \end{equation}
Putting $Z=|Z|e^{i\phi}$ where $\phi\in[-\pi,\pi)$, $h_1(Z)$ covers the plane
minus a wedge, e.g. $\arg h_1(Z)\in[-(1-\delta)\pi,(1-\delta)\pi]$.
This represents Minkowski space with a deficit angle $2\pi\delta$ and
may be considered to describe a cosmic string in the region outside the
spherical wave as shown in Fig.~\ref{Fig.2}.
(Alternatively, if $\delta<0$ and the exterior region may be taken to be
complete and the range of $\phi$ reduced so that the deficit angle representing
the string appears inside the sphere.)
Outside the spherical wavefront, the string may be taken to be located
along the axis $\eta=0$. Then, by putting $\eta=|\eta|e^{i\phi}$, it may be
observed from (\ref{transe3}) that, for the particular function $h_1(Z)$,
$\arg\eta=\arg h_1$. Thus, {\it the deficit angle is constant} along the
length of the string, corresponding to a constant tension.

For the case when $h=h_1(Z)$, the metric function $H_1(Z)$ in (\ref{L=0en0})
is given by
 \begin{equation}
 H_1={{1\over2}\delta(1-{1\over2}\delta)\over Z^2}.
 \label{H1}
 \end{equation}
 This indicates a singular point at $Z=0$, which is located at the South pole
of the spherical wavefront. However, there is an additional singularity
at $V=0$ which, for $\epsilon=0$, is located at the North pole at any time
$t>0$. As suggested at the end of section II, a much better physical
interpretation can be given to the $\epsilon\ne0$ solutions of the form
(\ref{L=0en0}). For $\epsilon=\pm1$, the physical singularity at $V=0$ appears
only in the vertex of the conical impulsive surface $U=0$ at $t=0$. Moreover,
according to (\ref{Psi4}) and (\ref{H1})
$$ \Phi_{22} \sim {(1+\epsilon Z\bar Z)^4\over(Z\bar Z)^2}\,
{U\,\delta(U)\over V^2}. $$
Therefore, there is a divergence not only at the South pole where $Z=0$
given by $\Phi_{22}\sim|Z|^{-4}$, but also at the North pole where $Z=\infty$
given by $\Phi_{22}\sim|Z|^4$. Both these divergences are equivalent and
indicate particles of {\it equal} ``mass'' at the two ends of the string.

\section{Lorentz transformations}

At this point, let us consider a space-time given by the line element
(\ref{L=0en0}) for some particular function $H$ and observe that it can best
be interpreted by first calculating the associated function $h(Z)$. This can
be achieved by integrating the three first order equations following
(\ref{U>0}). However, the resulting function is not unique as it contains
three complex constants of integration. In fact, this arbitrariness
corresponds to the freedom associated with the Lorentz transformations
 $$ Z\to h(Z)= {aZ+b\over cZ+d}, $$
 where $a,b,c,d$ are arbitrary complex constants satisfying the complex
condition $ad-bc=1$. It is well known that this is the most general {\it
global} holomorphic (i.e. analytic and conformal) transformation of the
Riemann sphere to itself (see chapter~1 of \cite{PenRin84}). It may also be
noted that the Schwarzian derivative (\ref{Schwarz}), i.e. $H(Z)$,
is invariant under this transformation.

In particular, a rotation about an arbitrary direction can be given by
 \begin{equation}
 h_r(Z)={Ze^{i\psi}\cos{\theta\over2}-\sin{\theta\over2}
\over Ze^{i\psi}\sin{\theta\over2}+\cos{\theta\over2}} \>e^{i\phi} ,
 \label{rotation}
 \end{equation}
 where $\theta,\phi,\psi$ are the Euler angles. We will adopt the following
construction: first we perform a rotation through $\psi$ about the $z$-axis,
then we rotate through $\theta$ about the original $y$-direction, and finally
we perform a rotation through $\phi$ about the original $z$-axis. This moves
the North pole to a point on the sphere given by the spherical coordinates
$\theta,\phi$.

Of the remaining Lorentz transformations, a boost in the $z$-direction
with velocity $v$ is given simply by
 \begin{equation}
h_b(Z)=wZ,
 \label{boost}
 \end{equation}
where $w=\sqrt{(1+v)/(1-v)}$,
which corresponds to a uniform expansion in the complex plane. It leaves
invariant both the poles of the Riemann sphere. Also, the null rotation
$h_n(Z)=Z-Z_0$, where $Z_0$ is complex, is a uniform translation in the
complex plane. It leaves invariant only the North pole of the Riemann
sphere. These are well described with useful pictures in chapter~1
of \cite{PenRin84}.

\section{The collision of cosmic strings}

By combining operations of the type given in (\ref{h1}), which introduces a
deficit angle, with suitable Lorentz transformations, we can generate much more
general solutions with predetermined physical properties. In particular, we
can use this technique to construct an explicit solution describing the
situation outlined by Nutku and Penrose~\cite{NutPen92} in which {\it two
cosmic strings collide and split} generating an impulsive spherical
gravitational wave.

It is assumed that two strings are initially moving with constant velocity and
approach each other. Of course, provided they are not parallel, it is always
possible to make a Lorentz transformation to a frame of reference in which the
two cosmic strings are orthogonal. There is therefore no loss of generality in
considering only orthogonal strings. It is also assumed that both strings snap
at their point of intersection and that the snapped ends propagate along the
length of the string at the speed of light. This will generate an impulsive
spherical gravitational wave.

We start with the simplest situation in which two orthogonal strings approach
each other with a negligible velocity. The solution for such a situation
can be constructed as follows.

Starting with (\ref{h1}), we first apply a rotation (\ref{rotation}) with
$\psi=\theta=\phi=\pi/2$. This leads to $(ih_1-1)/(h_1-i)$, which represents a
cosmic string located along the $y$-axis. Then we introduce another string in
the same way as in (\ref{h1}) by taking the $(1-\varepsilon)$-th power. This
removes a wedge which represents a second string that is located along the
$z$-axis and has deficit angle $2\pi\varepsilon$. The final mapping is
given by
\begin{equation}
 h_2(Z)=\left({iZ^{1-\delta}-1\over Z^{1-\delta}-i}\right)
 ^{1-\varepsilon}.
\label{h2}
\end{equation}
 This solution describes two snapped orthogonal cosmic strings
which were  at ``relative rest'' initially. The corresponding
function $H_2(Z)$ has the form
\begin{equation}
 H_2(Z)={{1\over2}\delta(1-{1\over2}\delta)\over Z^2}
-{{1\over2}\varepsilon(1-{1\over2}\varepsilon)4(1-\delta)^2Z^{-2\delta}
\over({Z^{1-\delta}+i)^2(Z^{1-\delta}-i})^2}.
\label{H2}
\end{equation}
The solution is visualized in Fig.~\ref{Fig.3}. On the left there is the
complex function $h_2(Z)$ in the complex Argand plane, on the right the
corresponding Riemann sphere obtained using the stereographic
identification. As can be seen, there are two perpendicular wedge
cuts on the sphere indicating that the flat space outside the spherical
impulsive gravitational wave contains two orthogonal pairs of cosmic
strings. This solution thus represents two cosmic strings which snapped at
the initial time $t=0$ and at their point of intersection.

We can now similarly construct a more general solution describing a situation
in which the two cosmic strings were in finite relative motion before the
collision. In this case, the four snapped half-strings are also in relative
translational motion. The more involved construction of such a solution is
shown in Figs.~\ref{Fig.4} and \ref{Fig.5}.

We start with the initial cut \ $h_1(Z)=Z^{1-\delta}$ \ given by (\ref{h1})
introducing the first string with deficit angle $2\pi\delta$ along the
$z$-axis as indicated in Fig.~\ref{Fig.2}. A subsequent rotation
(\ref{rotation}) with $\theta=\phi=\pi/2$, $\psi=\pi$ places the initial poles
$Z=0$ and $Z=\infty$ on the $y$-axis with the cut passing through the South
pole. The corresponding function \ $h_a(Z)=i(h_1+1)/(h_1-1)$ \ is
illustrated in
the complex Argand plane in Fig.~\ref{Fig.4}~(a) and on the Riemann sphere in
Fig.~\ref{Fig.5}~(a). Next we make a boost (\ref{boost}) with $w=w_1$ in the
$z$-direction as shown in Fig.~\ref{Fig.4}~(b) and Fig.~\ref{Fig.5}~(b).
Then another rotation (\ref{rotation}) with $\theta=-\pi/2$, $\phi=\psi=0$
about the $y$-axis leads to
 \begin{equation}
h_c(Z)=-{(w_1-i)Z^{1-\delta}+(w_1+i)\over(w_1+i)Z^{1-\delta}+(w_1-i)},
 \label{hc}
 \end{equation}
and puts the poles and the cut along the $z=0$ plane, symmetric
about the $x$-axis, see Fig.~\ref{Fig.4}~(c) and Fig.~\ref{Fig.5}~(c). At this
point we make a second cut from the North to South poles through the negative
$x$-axis by taking $h_c^{1-\varepsilon}$. This introduces the second string
with deficit angle $2\pi\varepsilon$ along the $z$-axis as indicated in
Fig.~\ref{Fig.4}~(d) and Fig.~\ref{Fig.5}~(d).
(Note that the function $h_c^{1-\varepsilon}$ for $w_1=1$ exactly reduces to
the previous form (\ref{h2}) with the poles located at
$h_c(0)=(i+w_1)/(i-w_1)=-i$ and $h_c(\infty)=\bar h_c(0)=i $.)
The next step is to make a rotation about the $y$-axis using $\theta=\pi/2$,
$\phi=\psi=0$ (see Fig.~\ref{Fig.4}~(e) and Fig.~\ref{Fig.5}~(e)). Finally
we perform a second boost $w_2$ in the negative $z$-direction. We thus
obtain
\begin{equation}
 h_2(Z)=w_2{h_c^{1-\varepsilon}-1\over h_c^{1-\varepsilon}+1}.
 \label{h2c}
 \end{equation}
This expression represents one cut through the North pole parallel to the
$x$-direction, and a second cut through the South pole parallel to the
$y$-direction, but with {\it both pairs of strings separated either side}
of the plane $z=0$ as shown in Fig.~\ref{Fig.4}~(f) and
Fig.~\ref{Fig.5}~(f). The resulting metric function using (\ref{h2c}) with
(\ref{hc}) is given by
 \begin{eqnarray}
 && H_2(Z)={{1\over2}\delta(1-{1\over2}\delta)\over Z^2}
-{{1\over2}\varepsilon(1-{1\over2}\varepsilon)16w_1^2(1-\delta)^2Z^{-2\delta}
\over [(w_1+i)Z^{1-\delta}+(w_1-i)]^2[(w_1-i)Z^{1-\delta}+(w_1+i)]^2}.
 \label{H2c}
 \end{eqnarray}
This is the explicit solution which describes the collision and consequent
snapping of two orthogonal cosmic strings as outlined by Nutku and Penrose
\cite{NutPen92}.

It may be noted that there exist null ``particles'' at the ends of the four
semi-infinite strings. These are
located on the spherical impulsive wave surface at points given by
$\xi_{1+}=h_2(h_c(Z=0))=iw_2\tan[(1-\varepsilon)(\phi_1-\pi/2)]$,
where $\phi_1=\arg(w_1+i)$, i.e. $\cot\phi_1=w_1$,
$\xi_{2+}=h_2(h_c(Z=\infty))=\bar\xi_{1+}$, and
$\xi_{3+}=h_2(h_c=0)=-w_2$, $\xi_{4+}=h_2(h_c=\infty)=w_2$. These
two pairs of points describe the four singular points in the
complex Argand plane of Fig.~\ref{Fig.4}~(f). Using the stereographic
relation to standard spherical coordinates,
$\xi=\cot(\theta/2)\,e^{i\phi}$, the corresponding four points on
the Riemann sphere --- the ends of the two cuts of
Fig.~\ref{Fig.5}~(f) --- have
$\cot(\theta_1/2)=|\xi_{1+}|=|\xi_{2+}|$ and
$\cot(\theta_2/2)=|\xi_{3+}|=|\xi_{4+}|$. The values of $\theta_1$
and $\theta_2$ can be made arbitrary by a suitable choice of the
boost parameters $w_1$ and $w_2$. Consequently, the cuts representing the
strings can be distributed arbitrarily over the spherical impulsive wave
(parallel to the $x$ and $y$-directions in our construction). It is natural to
consider a geometrically privileged situation in which the two cuts are
distributed {\it symmetrically}, as indicted in Fig.~\ref{Fig.5}~(f).
Obviously, this is given by the condition $\theta_1+\theta_2=\pi$ implying
$\cot(\theta_1/2)=\tan(\theta_2/2)$, i.e. $|\xi_{1+}||\xi_{3+}|=1$. Therefore,
the symmetry condition can be expressed as
 \begin{equation}
 w_2=\sqrt{\cot[(1-\varepsilon)(\pi/2-\phi_1)]},\
 \hbox{where}\ \cot\phi_1=w_1.
 \label{sym}
 \end{equation}
 For a negligible $\varepsilon$, $w_2\approx1/\sqrt{w_1}$.

Note that the above choice of the boost parameters in our construction results
in a geometrically symmetric situation in which the two semi-infinite parts of
the first snapped cosmic string are parallel to the $y$-axis and propagate in
the positive $z$-direction with speed $\cos\theta_1\equiv |v_2|$. Also, the
two semi-infinite parts of the second string are parallel to the $x$-axis and
move in the negative $z$-direction with the same speed $-|v_2|$. For
$\varepsilon=\delta$ the strings have equal tension, and the geometrically
symmetric situation corresponds to a choice of physically privileged
coordinates in which the geometrical origin of the impulse coincides with the
centre of mass of the colliding system.

Finally, it can be observed that the solution describing the collision
and snapping of two cosmic strings can alternatively and equivalently be
constructed if, starting again with (\ref{h1}), we perform the rotation
(\ref{rotation}) using $\theta=\phi=\pi/2$, $\psi=0$ in step (a), and the
rotation $\theta=\pi/2$, $\phi=\psi=0$ in step (c). Taking the other four
steps to be the same as in the previous construction shown in
Fig.~\ref{Fig.4} and Fig.~\ref{Fig.5}, this leads to
\begin{equation}
h_c(Z)={(w_1+i)Z^{1-\delta}-(w_1-i)\over(w_1-i)Z^{1-\delta}-(w_1+i)}.
 \label{hcalt}
 \end{equation}
This results in the solution given by (\ref{h2c}) with (\ref{hcalt})
presented in Fig.~\ref{Fig.6}. Again, there is a cut through the North
pole, parallel to
the $x$-direction and a second cut through the South pole, parallel to
the $y$-direction, but these cuts are {\it smaller} than in
Fig.~\ref{Fig.5}~(f). The resulting metric function is given by
 \begin{eqnarray}
 H_2(Z)={{1\over2}\delta(1-{1\over2}\delta)\over Z^2}
-{{1\over2}\varepsilon(1-{1\over2}\varepsilon)16w_1^2(1-\delta)^2Z^{-2\delta}
\over [(w_1+i)Z^{1-\delta}-(w_1-i)]^2[(w_1-i)Z^{1-\delta}-(w_1+i)]^2}.
 \label{H2calt}
 \end{eqnarray}
The two cuts representing the strings are distributed symmetrically
along the spherical impulsive wave if
$w_2=\sqrt{\cot[(1-\varepsilon)\phi_1]}$.

\section{Conclusions}

We have described the Penrose ``cut and paste'' method in detail and full
generality, emphasizing its geometrical properties and physical
interpretation. Using this, we have explicitly constructed the exact solution
which describes two colliding cosmic strings which snap at the point of their
intersection. The ends of each semi-infinite string are singular points
(interpreted as null particles) located on a sphere which is expanding
with the speed of light. This generates a spherical impulsive
gravitational wave.

\section*{Acknowledgments}

This work was supported by a visiting fellowship from the Royal Society and,
in part, by the grant GACR-202/99/0261 of the Czech Republic.

\newpage

\begin{figure}
\caption{
The stereographic correspondence between the Riemann sphere and the
complex Argand plane enables a geometrical description of the
Penrose junction conditions. Mapping in the complex plane $Z\to h(Z)$ is
equivalent to mapping points $P_-$ inside the impulsive spherical
surface to the corresponding points $P_+$ outside.}
\label{Fig.1}
\end{figure}

\begin{figure}
\caption{
The complex mapping $Z\to h_1(Z)=Z^{1-\delta}$ (left)
corresponds
to the cut through the Riemann sphere (right). This
represents
Minkowski space with a deficit angle $2\pi\delta$ outside
the
expanding spherical impulse, i.e. a ``snapping cosmic string''
along
the $z$-axis.}
\label{Fig.2}
\end{figure}

\begin{figure}
\caption{
The
mapping $Z\to h_2(Z)$ given by Eq.~(19) in the complex Argand
plane (left)
corresponds to two orthogonal cuts through
the Riemann sphere (right). This
describes two cosmic strings
along the $y$ and $z$-axes which are
``snapped'' at their
point of
intersection.}
\label{Fig.3}
\end{figure}

\begin{figure}
\caption{
Representation in the complex Argand plane of the construction of
the solution for two
cosmic strings parallel to the $x$ and
$y$-axes and moving apart in the
positive and
negative
$z$-directions.}
\label{Fig.4}
\end{figure}

\begin{figure}
\caption{
Representation on the Riemann sphere of the construction of the
solution for
two cosmic strings parallel to the $x$ and $y$-axes
and moving apart in the
positive and negative
$z$-directions.}
\label{Fig.5}
\end{figure}

\begin{figure}
\caption{
An alternative construction of the solution for two cosmic strings
parallel to
the $x$ and $y$-axes. The mapping $Z\to h_2(Z)$
given by Eqs.~(22) and (25)
in the complex Argand plane (left) corresponds
to two cuts through the
Riemann sphere (right) which are smaller
than those in
Fig.~5.}
\label{Fig.6}
\end{figure}


\begin{thebibliography}{99}


\bibitem{GlePul89} Gleiser R and Pullin J 1989
{\it Class. Quantum Grav.} {\bf 6} L141

\bibitem{BicSch89} Bi\v{c}\'ak J and Schmidt B 1989
{\it Class. Quantum Grav.} {\bf 6} 1547

\bibitem{Bicak90}  Bi\v{c}\'ak J 1990 {\it Astron. Nachr.} {\bf 311} 189

\bibitem{Pen72} Penrose R 1972 {\it General Relativity} ed
 L~O'Raifeartaigh (Oxford: Clarendon)

\bibitem{NutPen92} Nutku Y and Penrose R 1992
{\it Twistor Newsletter} No. 34, 11 May, 9

\bibitem{Hogan93} Hogan P A 1993 {\it Phys. Rev. Lett.} {\bf 70} 117

\bibitem{Hogan94} Hogan P A 1994 {\it Phys. Rev.} D {\bf 49} 6521

\bibitem{Hogan92} Hogan P A 1992 {\it Phys. Lett.} A {\bf 171} 21

\bibitem{PodGri99} Podolsk\'y J and Griffiths J B 1999
{\it Class. Quantum Grav.} {\bf 16} 2937

\bibitem{Hort90} Horta\c{c}su M 1990 {\it Class. Quantum Grav.} {\bf 7} L165

\bibitem{Hort99} Horta\c{c}su M 1996 {\it Class. Quantum Grav.} {\bf 13} 2683

\bibitem{BiHoOz96} Bilge A H, Horta\c{c}su H and \"Ozdemir N 1996
 {\it Gen. Relativ. Gravit.} {\bf 28} 511

\bibitem{Hogan95} Hogan P A 1995 {\it Lett. Math. Phys.} {\bf 35} 277

\bibitem{PenRin84} Penrose R and Rindler W 1984
 {\it Spinors and Space-time, vol.1}
 (Cambridge: Cambridge University Press)


\end{thebibliography}
\end{document}